\documentclass[12pt]{article}
\usepackage{epsfig,a4}

 \advance\oddsidemargin by -1in
 \advance\evensidemargin
 by -1in
 \marginparsep
 0.4cm
 \advance\topmargin by
 -0.0in
 \makeindex

 \newcommand\la{\langle}
 \newcommand\ra{\rangle}
 \newcommand\beq{\begin{equation}}
 
 \newcommand\eeq{\end{equation}}
 \newcommand\beqn{\begin{eqnarray}}
 \newcommand\eeqn{\end{eqnarray}}
 \newcommand\GeV{{\rm GeV}}

\def\BA{\begin{eqnarray}}
\def\BE{\begin{equation}}
\def\BF{\begin{figure}[htb]}
\def\BT{\begin{table}[htb]}
\def\EA{\end{eqnarray}}
\def\EE{\end{equation}}
\def\EF{\end{figure}}
\def\ET{\end{table}}

\def\la{\langle}
\def\ra{\rangle}

\def\fm{\,\mbox{fm}}
\def\GeV{\,\mbox{GeV}}

\def\lsim{\mathrel{\rlap{\lower4pt\hbox{\hskip1pt$\sim$}}
    \raise1pt\hbox{$<$}}}         
\def\gsim{\mathrel{\rlap{\lower4pt\hbox{\hskip1pt$\sim$}}
    \raise1pt\hbox{$>$}}}         

\begin{document}


\begin{center}
{\LARGE \bf
Color Transparency at Low Energies:
\\
\vspace*{0.2cm}
Predictions for JLAB
}
\end{center}

\begin{center}

\vspace{0.5cm}
 {\large B.Z.~Kopeliovich$^{1,2}$, J.~Nemchik$^{3}$ and Ivan~Schmidt$^{1}$}
 \\[1cm]
 {$^{1}$\sl Departamento de F\'{\i}sica y Centro de Estudios
Subat\'omicos,\\ Universidad T\'ecnica Federico Santa Mar\'{\i}a,
Valpara\'{\i}so, Chile} \\
 {$^{2}$\sl Joint Institute for Nuclear Research, Dubna, Russia} \\
 {$^{3}$\sl
Institute of Experimental Physics SAS, Watsonova 47,
04001 Kosice, Slovakia}

\end{center}

\vspace{1cm}

\date{\today}

\begin{abstract}

The study of color transparency (CT) in elastic electroproduction of
vector mesons off nuclei encounters the problem of the onset of
coherence length (CL) effects. The problem of CT-CL separation
arises especially at medium energies, corresponding to HERMES
experiment, when the coherence length is of the order of the nuclear
radius $R_A$. Only at asymptotic large energies, corresponding to
large CL, $l_c\gg R_A$, the CT-CL mixing can be eliminated. On the
other hand, the net CT effects can be studied in the kinematic
range accessible by the CLAS experiment, since in this case the CL
is much smaller than the nuclear radius. Using light-cone quantum
chromodynamics (QCD) dipole formalism we investigate manifestations
of CT effects in electroproduction of vector mesons. Motivated by
expected data from the CLAS experiment at JLab, we predict the $A$
and $Q^2$ dependence of nuclear transparency for $\rho^0$ mesons
produced incoherently off nuclei. We demonstrate that in the CLAS
kinematic region the CL effects are weak enough to keep the photon
energy at such values as to obtain maximal photon virtualities
keeping optimal statistics of the data. This has a clear advantage
in comparison with a standard investigation of net CT effects fixing
CL.

\end{abstract}



%
\section{Introduction}
\label{intro}
%

One of the fundamental phenomena coming from quantum chromodynamics
(QCD) is color transparency (CT), studied intensively  for more than
two decades. CT was predicted to occur already in 1981
\cite{zkl,bbgg}, in diffractive interaction with nuclei. The CT
phenomenon is manifested as a vanishing interaction cross section
$\sigma_{\bar qq}(r)\propto r^2$ \cite{zkl} for vanishing hadron
(quark configuration) transverse size $r$. As a result the nuclear
medium is more transparent for smaller transverse size of the hadron
(quark configurations). The history of CT investigation was
continued further in Refs. \cite{m-82,b-82}, where CT was predicted
to be manifested also in quasielastic high-$p_T$ scattering of
electrons and hadrons off nuclei with high momentum transfer. Later,
CT was predicted to appear also in quasi-free charge-exchange
scattering \cite{zk-87,preasymp}. It was also proposed \cite{k,fms}
to search for strong CT effects in reaction of 
back-to-back dijet production in coherent diffractive 
dissociation on 
nuclei.

Searches of a CT signal in quasielastic electron scattering
$A(e,e'p)A^*$ have not been successful so far. The first E18
experiment at SLAC \cite{slac} and later measurements at Jefferson Lab
\cite{e91013,e94139} failed to detect any signal of CT, which was
expected to be a deviation from the Glauber model calculations (see
the recent overview in \cite{kawtar}). The energy of recoil protons
was too low, and the formation length was too short in these
experiments. A promising way of detecting a signal of CT at these
energies would be a measurement of an asymmetric dependence of nuclear
transparency on missed energy \cite{jk}.

Searches of CT effects in quasielastic proton scattering, 
$A(p,2p)A^*$, were not conclusive either. Although a very interesting
variation of nuclear transparency was observed \cite{bnl},
no unambiguous interpretation has been proposed so far.
The observed effect maybe related either to the energy threshold for 
charm production \cite{stan}, or to interplay between hard an soft 
components of the production mechanism \cite{p-r} (see, however, 
finite energy corrections in \cite{preasymp}). One can find more 
details in the review \cite{review}.

Only a few experiments were able to confirm the CT
phenomenon. A very strong signal was observed in the PROZA
experiment at Serpukhov \cite{proza-87}, in quasi-free
charge-exchange pion scattering of nuclei, at an energy of
$40\,\GeV$. A sizable CT effect was detected in the E791 experiment 
\cite{e791} in diffractive coherent 
dissociation of pions on nuclei.

An observation of the onset of CT in virtual diffractive
photoproduction of $\rho$ mesons was claimed by the E665 collaboration
\cite{e665-rho}, measuring the $Q^2$-dependence of nuclear
transparency defined as: 
 \beq
Tr_A^{inc} = \frac{\sigma_{\gamma^*A\rightarrow VX}^{inc}}
                  {A~\sigma_{\gamma^*N\rightarrow VX}}
\label{15}
\eeq
 %
for the diffractive incoherent (quasielasic) production of vector
mesons, $\gamma^*A\rightarrow VX$\footnote{ An analogous definition
of nuclear transparency can also be done for the coherent or elastic
process, $\gamma^*A\rightarrow VA$.}.

The observed signal of CT by the E665 collaboration \cite{e665-rho}
had been predicted in \cite{knnz-93,knnz-94} as a rising nuclear
transparency (vanishing final state interaction) with increasing
hardness of the reaction $Q^2$. Although electroproduction of vector
mesons off nuclei is a very effective tool for the study of CT,
available experimental setups do not allow to reach such kinematic
region where a strong signal of CT is expected. In these cases one
has to deal only with an onset of this phenomenon. Therefore for a
detailed study of CT other effects which compete with the phenomenon
of CT should be considered. This was first done in
Refs.\cite{kn-95,hkn-96} within the Glauber model. There it was
shown that the main competition to CT effects at moderate energies
is the appearance of the so called coherence length (CL) effects,
leading to an analogous $Q^2$-dependence of nuclear transparency.
Later, in 1997, a multichannel evolution equation for the density
matrix was developed \cite{hk-97}, describing the propagation of a
hadronic wave packet through nuclear matter. This approach, applied
to electroproduction of vector mesons off nuclei in the hadronic
basis, incorporates both CT and CL effects. It was also suggested in
this reference to separate the net CT signal by investigating the
$Q^2$ dependence of nuclear transparency at values of the photon
energy $\nu$ which keep the CL constant.

An alternative description of CT in electroproduction of vector
mesons can be realized also within the quark-gluon representation
\cite{kz-91,knnz-93,knnz-94,jan97,knst-01,n-02,n-03}. Here a photon
of high virtuality $Q^2$ is expected to produce a pair with a small
$\sim 1/Q^2$ transverse separation\footnote{In fact, the situation
is somewhat more complicated. For very asymmetric pairs the $q$ or
$\bar q$ carry almost the whole photon momentum, and then the pair
can have a large separation, see for example Ref.~\cite{knst-01}}.
Then CT manifests itself as a vanishing absorption of the small size
colorless $\bar qq$ wave packet during propagation through the
nucleus. The dynamical evolution of this small size $\bar qq$ pair
to a normal size vector meson is controlled by the time scale called
formation time. Due to uncertainty principle, one needs a time
interval to resolve different levels $V$ (the ground state) or $V'$
(the next excited state) in the final state. In the rest frame of
the nucleus this formation time is Lorentz dilated,
%
 \beq
t_f = \frac{2\,\nu}
{\left.m_{V^\prime}\right.^2 - m_V^2}\ .
\label{20}
 \eeq
%
A rigorous quantum-mechanical description of the pair evolution was
suggested in \cite{kz-91}, based on the nonrelativistic light-cone
(LC) Green function technique. The same LC Green function formalism
has been already applied also for Drell-Yan production in
proton-nucleus and nucleus-nucleus interactions \cite{krtj-03}, for
nuclear shadowing in DIS \cite{krt-00,dis-num}, and for coherent and
incoherent electroproduction of vector mesons off nuclei
\cite{knst-01,n-02,n-03}.

Another phenomenon known to cause nuclear suppression is quantum
coherence. It results from destructive interference of the
amplitudes for which the interaction takes place on different bound
nucleons. It is characterized by the coherence length (CL), which is
related to the longitudinal momentum transfer by $q_c=1/l_c$. It
also corresponds to the coherence time ($l_c=t_c$), given by

%
 \beq
t_c = \frac{2\,\nu}{Q^2 + m_V^2}\ .
\label{30}
 \eeq
%

In electroproduction of vector mesons off nuclei one needs to
disentangle CT (absorption) and CL (shadowing) as the two sources of
nuclear suppression. Conventionally, one can associate these effects
with final and initial state interactions, respectively. Detailed
analysis of the CT and CL effects in electroproduction of light
vector mesons off nuclei showed \cite{knst-01} that the coherence
length is larger or comparable with the formation one, $l_c\gsim
l_f$, starting from the photoproduction limit up to $Q^2 \sim (1\div
2)\GeV^2$. This does not happen in charmonium production, where
there is a strong inequality $l_c < l_f$, independent of $Q^2$ and
$\nu$ \cite{n-02,n-03}, and which leads to a different scenario of
CT-CL mixing in comparison with the production of light vector
mesons.

Recently new HERMES data \cite{hermes-a1,hermes-a2} have been
gradually obtained for diffractive exclusive electroproduction of
$\rho^0$ mesons on nitrogen target. At the beginning the data were
presented as a dependence of nuclear transparency (\ref{15}) on
coherence length (\ref{30}). The data for incoherent $\rho^0$
production decrease with $l_c$, as expected from the effects of
initial state interactions. On the other hand, the nuclear
transparency for coherent $\rho^0$ production increases with
coherence length, as expected from the effects of the nuclear form
factor \cite{knst-01}. However, each $l_c$- bin of the data contains
different values of $\nu$ and $Q^2$, i.e. there are different
contributions of both effects, CT and CL. For this reason, the
$l_c$- behavior of nuclear transparency does not allow to study
separately CT and CL effects. Therefore it was proposed in
\cite{hk-97,knst-01} that CT can be separately studied eliminating
the effect of CL from the data on the $Q^2$ dependence of nuclear
transparency, in a way which keeps $l_c = const$. According to this
prescription, later the HERMES data \cite{hermes-a2} were presented
as $Q^2$ dependence of nuclear transparency at different fixed
values of $l_c$. Then the rise of $Tr_A$ with $Q^2$ represents a
signature of CT. The HERMES data \cite{hermes-a2} are in a good
agreement with the predictions coming from Ref.~\cite{knst-01}. New
HERMES data on neon and krypton target should be available soon.
This will allow to verify further the predictions for CT from
Ref.~\cite{knst-01}.

Moreover, another investigation of CT has been carried out by the
CLAS collaboration at JLab \cite{kawtar}, studying incoherent
electroproduction of $\rho$ mesons off nuclei at small photon
energies $2.2 < \nu < 4.5\,\GeV$. At such values of $\nu$ the
corresponding coherence length $l_c\ll R_A$, where $R_A$ is the
nuclear radius. For this reason CL effects are expected to be much
weaker than CT and CL-CT mixing does not play an important role.
Therefore the study of vector meson electroproduction at small
energies represents an alternative way for investigating a clear
signal of CT. Because new data from the CLAS collaboration will
appear soon we will present here the corresponding predictions for
CT. The main emphasis will be devoted to discussing the advantages
in the investigation of CT effects in the CLAS energy range in
comparison with the HERMES experiment.

The paper is organized as follows. In Sect.~\ref{lcc} we present a
brief summary of the light-cone dipole approach to diffractive
electroproduction of vector mesons. In the next Sect.~\ref{class} we
calculate the predictions for the $Q^2$ dependence of nuclear
transparency, for various nuclear targets, at the small energies
corresponding to the CLAS experiment at JLab. We analyze also the
strength of CL effects and their influence on the CT signal. We show
that CL-CT mixing at small energies does not spoil the clear onset
of CT effects and permits to treat the CLAS experimental values at
different small $l_c\ll R_A$, allowing to process the data at
maximal possible statistics. We discuss the fact that the weak CL
effects in the CLAS energy region lead also to the possibility of
studying the CT phenomenon at much larger values of $Q^2$ than those
which are presently available, using a prescription to fix CL, in
particular for the HERMES experiment. The results of the paper are
summarized and discussed in Sect.~\ref{conclusions}.

%
%
\section{A Short Review of the Color Dipole Phenomenology}
\label{lcc}
%
%

The light-cone (LC) dipole approach for the process $\gamma^{*}N\to
V~N$ was already used in Ref.~\cite{kz-91,hikt-00} to study the exclusive
photo- and electroproduction of charmonia, and in
Ref.~\cite{knst-01} for elastic virtual photoproduction of light
vector mesons $\rho^0$ and $\Phi^0$ (for a review see also
\cite{ins-05}). Therefore, we present only a short review of this
approach. Here a diffractive process is treated as elastic
scattering of a $\bar qq$ fluctuation of the incident particle.  The
elastic amplitude is given by a convolution of the universal flavor
independent dipole cross section for the $\bar qq$ interaction with
a nucleon, $\sigma_{\bar qq}$ \cite{zkl}, and the initial and final
wave functions. Then the forward production amplitude for the
process $\gamma^{*}N\to V~N$ can be represented in the
quantum-mechanical form
%
 \BE
{\cal M}_{\gamma^*N\rightarrow VN}(s,Q^{2})\, =\,
\langle\, V |\,\sigma_{\bar qq}({\vec{r}},s)\,|\gamma^{*}\,\rangle\, =\,
\int\limits_{0}^{1} d\alpha \int d^{2}{{r}}\,\,
\Psi_{V}^{*}({\vec{r}},\alpha)\,
\sigma_{\bar qq}({\vec{r}},s)\,
\Psi_{\gamma^{*}}({\vec{r}},\alpha,Q^2)\,
\label{120}
 \EE
%
 with the normalization
%
 \beq
\left.\frac{d\sigma(\gamma^*N\to VN)}{dt}\right|_{t=0} =
\frac{|{\cal M}_{\gamma^*N\to VN}(s,Q^2)|^{2}}{16\,\pi}.
\label{125}
 \eeq
%

There are three ingredients contributing to the amplitude
(\ref{120}):

(i) The dipole cross section $\sigma_{\bar qq}({\vec{r}},s)$ which
depends on the $\bar qq$ transverse separation $\vec{r}$ and the
c.m. energy squared $s$.

(ii) The LC  wave function of the photon,
$\Psi_{\gamma^{*}}({\vec{r}},\alpha,Q^2)$, which besides the
$\vec{r}$ dependence, depends also on the photon virtuality $Q^2$
and the relative share $\alpha$ of the photon momentum carried by
the quark.

(iii) The LC wave function $\Psi_{V}({\vec{r}},\alpha)$ of the
vector meson.

Detailed description of these ingredients can be found in Refs.
\cite{knst-01,n-02}.

Notice that in the LC formalism the photon and meson wave functions
contain also higher Fock states $|\bar qq\ra$, $|\bar qqG\ra$,
$|\bar qq2G\ra$, etc. The effect of these higher Fock states is
implicitly incorporated into the energy dependence of the dipole
cross section $\sigma_{\bar qq}(\vec{r},s)$, as is given in
Eq.~(\ref{120}).

Since our main interest is electroproduction of light vector mesons,
we explicitly consider the nonperturbative interaction effects
between the $q$ and $\bar q$, which can be included in the LC wave
function of the photon, $\Psi_{\gamma^{*}}({\vec{r}},\alpha,Q^2)$.
For this purpose we use the corresponding phenomenology based on the
LC Green function approach, developed in Ref.~\cite{kz-91,kst2}. The Green
function $G_{\bar qq}(z_1,\vec r_1;z_2,\vec r_2)$ describes the
propagation of an interacting $\bar qq$ pair between points with
longitudinal coordinates $z_{1}$ and $z_{2}$ and with initial and
final transverse separations $\vec r_1$ and $\vec r_2$. This Green
function satisfies the two-dimensional Schr\"odinger equation,
%
\BA
i\frac{d}{dz_2}\,G_{\bar qq}(z_1,\vec r_1;z_2,\vec r_2)
=\,
\left\{\frac{\epsilon^{2} - \Delta_{r_{2}}}{2\,\nu\,\alpha\,(1-\alpha)}
+ V_{\bar qq}(z_2,\vec r_2,\alpha)\,\right\}
G_{\bar qq}(z_1,\vec r_1;z_2,\vec r_2)\ .
\label{250}
\EA
%
Here the Laplacian $\Delta_{r}$ acts on the coordinate $r$.

The imaginary part of the LC potential $V_{\bar qq}(z_2,\vec
r_2,\alpha)$ in (\ref{250}) is responsible for the attenuation of
the $\bar qq$ pair in the medium, while the real part represents the
interaction between the $q$ and $\bar{q}$. This potential is
supposed to provide the correct LC wave functions of vector mesons.
For the sake of simplicity we use the oscillator form of the
potential,
%
 \BE
{\rm Re}\,V_{\bar qq}(z_2,\vec r_{2},\alpha) =
\frac{a^4(\alpha)\,\vec r_{2}\,^2}
{2\,\nu\,\alpha(1-\alpha)}\ ,
\label{260}
 \EE
%
 which leads to a Gaussian $r$-dependence of the LC wave function of the
meson ground state.  The shape of the function $a(\alpha)$ can be
found in Ref.~\cite{kst2}.

 In this case equation (\ref{250}) has an analytical solution leading
to an explicit form of the harmonic oscillator Green function
\cite{fg},
%
 \BA
G_{\bar qq}(z_1,\vec r_1;z_2,\vec r_2) =
\frac{a^2(\alpha)}{2\;\pi\;i\;
{\rm sin}(\omega\,\Delta z)}\, {\rm exp}
\left\{\frac{i\,a^2(\alpha)}{{\rm sin}(\omega\,\Delta z)}\,
\Bigl[(r_1^2+r_2^2)\,{\rm cos}(\omega \;\Delta z) -
2\;\vec r_1\cdot\vec r_2\Bigr]\right\}
\nonumber\\ \times
{\rm exp}\left[-
\frac{i\,\epsilon^{2}\,\Delta z}
{2\,\nu\,\alpha\,(1-\alpha)}\right] \ ,
\label{270}
 \EA
%
where $\Delta z=z_2-z_1$ and
%
 \BE \omega = \frac{a^2(\alpha)}{\nu\;\alpha(1-\alpha)}\ .
\label{280}
 \EE
%
 The boundary condition is $G_{\bar
qq}(z_1,\vec r_1;z_2,\vec r_2)|_{z_2=z_1}=
\delta^2(\vec r_1-\vec r_2)$.

Using concrete forms of all the above ingredients (specified in
Ref.~\cite{knst-01}, for example) we can calculate the forward
production amplitude for the process $\gamma^*\,N \to V\,N$,
separately for transverse and longitudinal photons and vector
mesons. Assuming $s$-channel helicity conservation (SCHC), the
forward scattering amplitude reads,
%
%
 \BA
{\cal M}_{\gamma^{*}N\rightarrow V\,N}^{T}(s,Q^{2})
\Bigr|_{t=0} &=&
N_{C}\,Z_{q}\,\sqrt{2\,\alpha_{em}}
\int d^{2} r\,\sigma_{\bar qq}(\vec r,s)
\int\limits_0^1 d\alpha \Bigl\{ m_{q}^{2}\,
\Phi_{0}(\epsilon,\vec r,\lambda)\Phi^T_{V}(\vec r,\alpha)
\nonumber\\
&-& \bigl [\alpha^{2} + (1-\alpha)^{2}\bigr ]\,
\vec{\Phi}_{1}(\epsilon,\vec r,\lambda)\cdot
\vec{\nabla}_{r}\,\Phi^T_{V}(\vec r,\alpha) \Bigr\}\,;
\label{360}
 \EA
%
 \BA
{\cal M}_{\gamma^{*}N\rightarrow V\,N}^{L}(s,Q^{2})
\Bigr|_{t=0} &=&
4\,N_{C}\,Z_{q}\,\sqrt{2\,\alpha_{em}}\,m_{V}\,Q\,
\int d^{2} r\,\sigma_{\bar qq}(\vec r,s)
\nonumber\\&\times&
\int\limits_0^1 d\alpha\,
\alpha^{2}\,(1-\alpha)^{2}\,
\Phi_{0}(\epsilon,\vec r,\lambda)
\Phi^L_{V}(\vec r,\alpha)\, ,
\label{370}
 \EA
%
%
with the normalization given by Eq.~(\ref{125}) for both T and L
polarizations. The functions $\Phi_{0,1}$ in Eqs.~(\ref{360}) and
(\ref{370}) include nonperturbative interaction effects between $q$
and $\bar q$, and $\Phi_{V}$ represents the vector meson wave
function, whose explicit form will be taken from
Ref.~\cite{knst-01}. The real part of the amplitude is included
according to the prescription described in Ref.~\cite{knst-01}. In
Eqs.~(\ref{360}) and (\ref{370}) the terms proportional to [$
\Phi_0(\epsilon,\vec r,\lambda)\,\Phi_V(\vec r,\alpha)$] and to [$
\vec{\Phi}_1(\epsilon,\vec r,\lambda)\cdot \vec{\nabla}_r
\Phi_V(\vec r,\alpha)$] correspond to the helicity conserving and
helicity-flip transitions in the $\gamma^*\to \bar qq$ and $V\to
\bar qq$ vertices, respectively. The helicity flip transitions
represent the relativistic corrections. For heavy quarkonium these
corrections become important only at large $Q^2\gg m_V^2$. For
production of light vector mesons, however, they are non-negligible
even in the photoproduction limit, $Q^2 = 0$.

Usually the data are presented in the form of the production cross
section $\sigma = \sigma^T + \epsilon\,'\,\sigma^L$, at fixed photon
polarization $\epsilon\,'$. Here the transverse and longitudinal cross 
sections, integrated over
$t$, read:
%
 \beq
\sigma^{T,L}(\gamma^{*}N\to VN) =
\frac{|{\cal M}^{T,L}|^{2}}
{16\pi\,B_{\gamma^*N}}\ ,
\label{375}
 \eeq
%
 where $B_{\gamma^*N}$ is the $t$-slope parameter in the differential
cross section for reaction $\gamma^*\,p \to V\,p$, Eq.~(\ref{125}). The
absolute value of the production cross section has already been checked in
Ref.~\cite{knst-01}, by comparing with data for elastic $\rho^0$ and
$\Phi^0$ electroproduction, and for charmonium exclusive electroproduction
$\gamma^*\,p \to J/\Psi\,p$ in Refs.~\cite{n-02,n-03}.

The generalization of the LC dipole approach to nuclear targets is
straightforward. We focus in the present paper on diffractive
incoherent (quasielastic) production of vector mesons off nuclei,
$\gamma^{*}\,A\rightarrow V\,X$, where the observable usually
studied experimentally is nuclear transparency, defined by
Eq.~(\ref{15}). Since the $t$-slope of the differential quasielastic
cross section is the same as on a nucleon target, instead of the
integrated cross sections one can also use the forward differential
cross sections, Eq.~(\ref{125}), to write
%
 \beq
Tr^{inc}_A = \frac{1}{A}\,
\left|\frac{{\cal M}_{\gamma^{*}A\to VX}(s,Q^{2})}
{{\cal M}_{\gamma^{*}N\to VN}(s,Q^{2})}\right|^2\, .
\label{485}
 \eeq
%

The nuclear forward production amplitude ${\cal M}_{\gamma^{*}\,A\to
V\,X}(s,Q^{2})$ in Eq.~(\ref{485}) is calculated using the LC Green
function approach \cite{knst-01}. In this approach the physical
photon $|\gamma^*\ra$ is decomposed into different Fock states,
namely, the bare photon $|\gamma^*\ra_0$, plus $|\bar qq\ra$, $|\bar
qqG\ra$, etc. As we mentioned above the higher Fock states
containing gluons describe the energy dependence of the
photoproduction reaction on a nucleon. Besides, these Fock
components also lead to gluon shadowing as far as nuclear effects
are concerned. Detailed description and calculation of gluon
shadowing for the case of vector meson production off nuclei is
presented in Refs.~\cite{knst-01,ikth-02}. However, in the CLAS and
HERMES kinematic range studied in the present paper the gluon
shadowing is negligible and therefore is not included in our
calculations.

The propagation of an interacting $\bar qq$ pair in a nuclear medium is also
described by the Green function satisfying the evolution Eq.~(\ref{250}).
However, the potential in this case acquires an imaginary part which
represents absorption in the medium,
%
 \BE
Im V_{\bar qq}(z_2,\vec r,\alpha) = -
\frac{\sigma_{\bar qq}(\vec r,s)}{2}\,\rho_{A}({b},z_2)\,,
\label{440}
 \EE
%
where $\rho_{A}({b},z)$ is the nuclear density function
defined at given impact parameter $b$ and longitudinal
coordinate $z$.

The analytical solution of Eq.~(\ref{250}) is only known for the
harmonic oscillator potential $V(r)\propto r^2$. To keep the
calculations reasonably simple we use the dipole cross section
approximation,
%
 \beq
\sigma_{\bar qq}(r,s) = C(s)\,r^2\ ,
\label{460}
 \eeq
%
which allows to obtain the Green function in an analytical form (see
Eq.~(\ref{270})). The energy dependent factor $C(s)$ was adjusted by
the procedure described in ref.~\cite{knst-01}.

With the potential that follows from Eqs.~(\ref{440}) -- (\ref{460})
the solution of Eq.~(\ref{250}) has the same form as
Eq.~(\ref{270}), except that one should replace $\omega \Rightarrow
\Omega$, where
%
 \beq
\Omega = \frac{\sqrt{a^4(\alpha)-
i\,\rho_{A}({b},z)\,
\nu\,\alpha\,(1-\alpha)\,C(s)}}
{\nu\;\alpha(1-\alpha)}\ .
\label{470}
 \eeq
%

The evolution equation (\ref{250}) was recently solved numerically
for the first time in Ref.~\cite{dis-num}, with the potential
$V_{\bar qq}(z_{2},\vec r_{2},\alpha)$ containing the imaginary part
(\ref{440}) and with the realistic dipole cross section given in
\cite{kst2}.

As is discussed in \cite{knst-01} the value of $l_c$ can distinguish
different regimes of vector meson production:

{\bf (i)} The region where CL is much shorter than the mean nucleon
spacing in a nucleus ($l_c \to 0$). In this case $G(z_2,\vec
r_2;z_1,\vec r_1) \to \delta(z_2-z_1)$. Correspondingly, the
formation time of the meson wave function is very short and is given
by Eq.~(\ref{20}). For light vector mesons $l_f\sim l_c$, and since
both the formation and coherence lengths are proportional to the
photon energy both must be short. Consequently, nuclear transparency
is given by the simple formula corresponding to the Glauber
approximation:
%
 \BA
Tr_A^{inc}\Bigr|_{l_c\ll R_A} &\equiv&
\frac{\sigma_V^{\gamma^*A}}
{A\,\sigma_V^{\gamma^*N}} =
\frac{1}{A}
\,\int d^2b\,
\int\limits_{-\infty}^{\infty}
dz\,\rho_A(b,z)\,
\exp\left[-\sigma^{VN}_{in}
\int\limits_z^{\infty} dz'\,
\rho_A(b,z')\right]\nonumber\\
&=&
\frac{1}{A\,\sigma^{VN}_{in}}\,
\int d^2b\,\left\{1 -
\exp\Bigl[-\sigma^{VN}_{in}\,T_A(b)\Bigr]\right\}=
\frac{\sigma^{VA}_{in}}{A\,\sigma^{VN}_{in}}\ ,
\label{490}
 \EA
%
where $\sigma_{in}^{VN}$ is
the inelastic $VN$ cross section.

{\bf (ii)} The production of charmonia and other heavy flavors
corresponds to the intermediate case where as before $l_c\to 0$, but
where $l_f\sim R_A$ can be realized. Then the formation of the meson
wave function is described by the Green function, and the numerator
of the nuclear transparency ratio Eq.~(\ref{485}) has the form
\cite{kz-91},
%
 \BA
\Bigl|{\cal M}_{\gamma^{*}A\to VX}(s,Q^{2})
\Bigr|^2_{l_c\to0;\,l_f\sim R_A} =
\int d^2b\int_{-\infty}^{\infty} dz\,\rho_A(b,z)\,
\Bigl|F_1(b,z)\Bigr|^2\ ,
\label{500}
 \EA
%
 where
%
 \BA
F_1(b,z) =
\int_0^1 d\alpha
\int d^{2} r_{1}\,d^{2} r_{2}\,
\Psi^{*}_{V}(\vec r_{2},\alpha)\,
G(z^\prime,\vec r_{2};z,\vec r_{1})\,
\sigma_{\bar qq}(r_{1},s)\,
\Psi_{\gamma^{*}}(\vec r_{1},
\alpha)\Bigl|_{z^\prime\to\infty}
\label{505}
 \EA
%

{\bf (iii)} In the high energy limit $l_c \gg R_A$. In this case
$G(z_2,\vec r_2;z_1,\vec r_1) \to \delta(\vec r_2 - \vec r_1)$, i.e.
all fluctuations of the transverse $\bar qq$ separation are
``frozen'' by Lorentz time dilation. Then the numerator on the
right-hand-side (r.h.s.) of Eq.~(\ref{485}) takes the form
\cite{kz-91},
%
 \BA
&&\Bigl|{\cal M}_{\gamma^{*}A\to VX}(s,Q^{2})
\Bigr|^2_{l_c \gg R_A}=
\int d^2b\,T_A(b)
\times
\nonumber\\
&&\left|\int d^2r\int_0^1 d\alpha \,\Psi_{V}^{*}(\vec r,\alpha)\,
\sigma_{\bar qq}(r,s)\,\times \exp\left[-{1\over2}\sigma_{\bar
qq}(r,s)\,T_A(b)\right] \Psi_{\gamma^{*}}(\vec
r,\alpha,Q^2)\right|^2,
\label{510}
 \EA
%
and the $\bar qq$ attenuates with a constant absorption cross
section as in the Glauber model, except that the whole exponential
is averaged rather than just the cross section in the exponent. The
difference between the results of the two prescriptions are the well
known inelastic corrections of Gribov \cite{zkl}.

{\bf (iv)} This regime reflects the general case when there are no
restrictions for either $l_c$ or $l_f$. The corresponding
theoretical tools have been developed only recently and for the
first time in \cite{knst-01}. In this general case the incoherent
nuclear production amplitude squared is represented as a sum of two
terms \cite{hkz},
%
 \BA
\Bigl|\,{\cal M}_{\gamma^{*}A\to
VX}(s,Q^{2})\Bigr|^{2} = \int d^{2}b
\int\limits_{-\infty}^{\infty} dz\,\rho_{A}({b},z)\,
\Bigl|F_{1}({b},z) - F_{2}({b},z)\Bigr|^{2}\ .
\label{520}
 \EA
%

 The first term $F_{1}({b},z)$, introduced above in Eq.~(\ref{505}),
corresponds to the short $l_c$ limit (ii). The second term
$F_{2}({b},z)$ in (\ref{520}) corresponds to the situation when the
incident photon produces a $\bar qq$ pair diffractively and
coherently at the point $z_1$, prior to an incoherent quasielastic
scattering at point $z$. The LC Green function describes the
evolution of the $\bar qq$ over the distance from $z_1$ to $z$ and
further on, up to the formation of the meson wave function.
Correspondingly, this term has the form,
%
 \BA
F_{2}(b,z) &=& \frac{1}{2}\,
\int\limits_{-\infty}^{z} dz_{1}\,\rho_{A}(b,z_1)\,
\int\limits_0^1 d\alpha\int d^2 r_1\,
d^2 r_{2}\,d^2 r\,
\times
\nonumber \\
&&\Psi^*_V (\vec r_2,\alpha)\,
G(z^{\prime}\to\infty,\vec r_2;z,\vec r)\,
\sigma_{\bar qq}(\vec r,s)\,
G(z,\vec r;z_1,\vec r_1)\,
\sigma_{\bar qq}(\vec r_1,s)\,
\Psi_{\gamma^{*}}(\vec r_1,\alpha)\, .
\label{530}
 \EA
%

Finally, we would like to emphasize that Eq.~(\ref{520}) correctly
reproduces the limits (i) - (iii) at small and large energies, as
was already analyzed in Ref.~\cite{knst-01}.

%
%
\section{Signal of CT
Expected at Low Energies}\label{class}
%
%

Exclusive incoherent electroproduction of vector mesons off nuclei
has been shown in \cite{knnz-94,knst-01} to be a very effective tool
for the investigation of CT. Increasing the photon virtuality $Q^2$
has the effect of squeezing the produced $\bar qq$ wave packet. Such
a small colorless system propagates through the nucleus with little
attenuation, provided that the energy is sufficiently high ($l_f\gg
R_A$), so the fluctuations of the $\bar qq$ separation are
``frozen'' during propagation. Consequently, a rise of the nuclear
transparency $Tr_A^{inc}(Q^2)$ with $Q^2$ should give a signal for
CT. Indeed, such a rise was observed in the E665 experiment
\cite{e665-rho} at Fermilab for exclusive production of $\rho^0$
mesons off nuclei, and this has been claimed as a manifestation of
CT.

However, the effect of coherence length \cite{kn-95,hkn-96} leads
also to a rise of $Tr_A^{inc}(Q^2)$ with $Q^2$ and so it can imitate
the CT effects. Both effects work in the same direction and this
produces the problem of CT-CL separation, which was solved in
Refs.~\cite{hk-97,knst-01}, where a simple prescription for the
elimination of CL effects from the data on the $Q^2$ dependence of
nuclear transparency was presented. One should bin the data in a way
which keeps $l_c = const$. It means that $\nu$ and $Q^2$ 
should be correlated, 
%
 \beq
\nu = {1\over2}\,l_c\,(Q^2+m_V^2)\ .
\label{534}
 \eeq
%
In this case any rise with $Q^2$ of the nuclear transparency ratio
is a clear signal of CT \cite{hk-97,knst-01}.

In the present paper we investigate a manifestation of CT effects in
the production of vector mesons at small energies, corresponding to
CLAS experiment at JLab. Motivated by expected new data from the
CLAS collaboration we concentrate on the production of $\rho^0$
vector mesons. Besides this experimental motivation, there is also
theoretical interest because the coherence and formation effects are
much more visible for the light than for the heavy vector mesons, as was
discussed and analyzed in Refs.~\cite{knst-01,n-02,n-03}.

We start, however, with a discussion about the first theoretical
analysis of CT phenomenon, related with the E665 experiment
\cite{e665-rho}. The corresponding predictions \cite{knnz-94} were
based on the assumption that the energy corresponding to the E665
experiment is high enough, so that  $l_f, l_c\gg R_A$, allowing to
use the ``frozen'' approximation (\ref{510}). Correspondingly one
can neglect a variation of CL with $Q^2$. However, this is true only
at $Q^2\lsim 3\div 5\,\GeV^2$, as was shown later in
Ref.~\cite{knst-01}. It means that fluctuations of the size of the
$\bar qq$ pair become important at larger $Q^2\gsim 5\,\GeV^2$, and
therefore at such values of $Q^2$ one should use the general
expression (\ref{520}) for the calculation of nuclear transparency.
Thus the observed variation of $Tr_A^{inc}(Q^2)$ is a net
manifestation of CT only at smaller $Q^2$. Although the model
predictions, within the color dipole approach \cite{knnz-94}, were
in agreement with the E665 data, small corrections for CL effects,
done at larger $5 < Q^2 < 10\,\GeV^2$ in Ref.~\cite{knst-01}, did
not spoil this agreement.

%
  \begin{figure}[bht]
\includegraphics{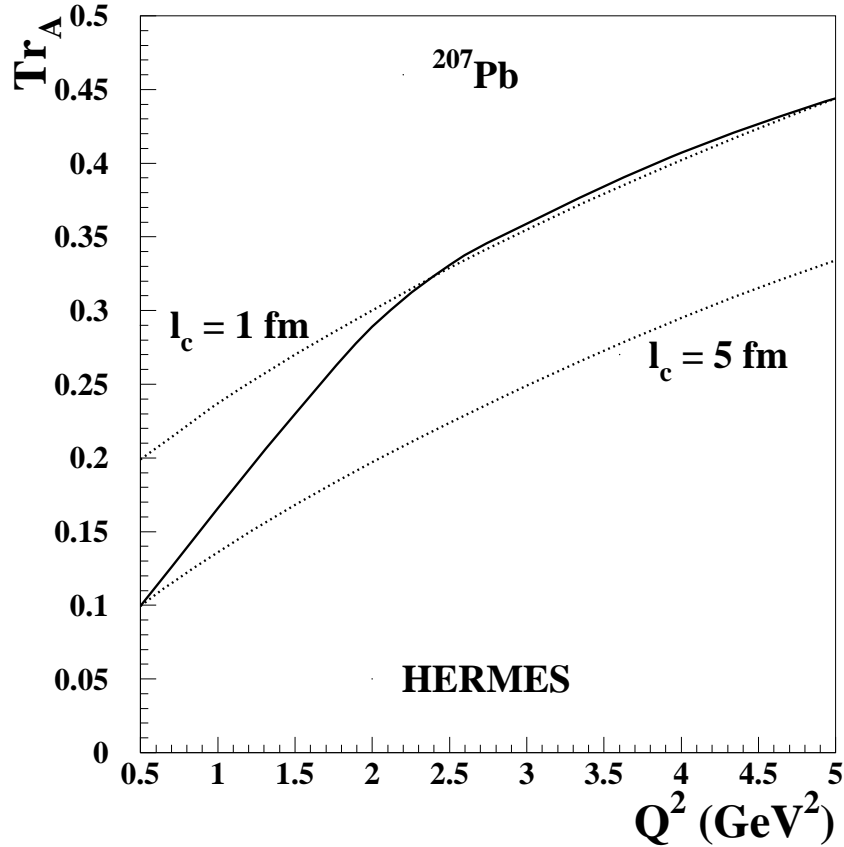}
\begin{center}
\vspace{11.0cm}
\parbox{13cm}
{\caption[Delta]
 {\sl $Q^2$ dependence of the nuclear transparency
$Tr_A^{inc}$ for exclusive electroproduction of $\rho$
mesons on $^{207}Pb$ target.
The dotted lines represent the CL fixed at $l_c = 1.0$
and $5.0\,\fm$. The solid lines represent predictions
at the fixed mean photon energy $\la\nu\ra = 13\,\GeV$
corresponding to HERMES experiment.
}
 \label{hermes}}
 \end{center}
 \end{figure}
%

On the other hand, CT effects are also under investigation at lower
energies in both the HERMES experiment at HERA and the CLAS
experiment at JLab, measuring the same process of incoherent
electroproduction of $\rho$ mesons. Because CL effects are also
important one should use the LC Green function formalism as a very
effective tool for such study, because then both CT and CL effects
are naturally incorporated. Due to a strong CL-CT mixing one should
eliminate the effect of CL according to the prescription mentioned
above (see Eq.~(\ref{534})). The corresponding predictions for CT
\cite{knst-01} are in a good agreement with the HERMES data
\cite{hermes-a2} on the $Q^2$ dependence of nuclear transparency at
different fixed values of $l_c$. Expected new HERMES data on neon
and krypton targets should allow to verify further the predictions
for CT from \cite{knst-01}.

Following the relation Eq.~(\ref{534}), CT effects are
found to be much stronger at low (HERMES, CLAS) than at high (E665)
energies \cite{knst-01}. At high energies the CL $l_c$ is long.
Consequently, the formation length is long too, $l_f\gsim l_c\gg
R_A$, and nuclear transparency rises with $Q^2$ only because the
mean transverse size of the $\bar qq$ photon fluctuations decreases.
It is not so at lower energies when $l_c \lsim R_A$. Then, according
to the prescription ~(\ref{534}), the photon energy rises with $Q^2$
and consequently the formation length, Eq.~(\ref{20}), rises as
well. Thus, these two effects add up leading to a steeper growth of
$Tr_A^{inc}(Q^2)$ for short $l_c$.

The prescription (\ref{534}) for studying net signals of CT fixing
the values of $l_c$, has one hidden disadvantage. Increasing $Q^2$
one needs to increase also the photon energy, and this can be done
only up to the kinematic limit given by the HERMES experiment,
$\nu_{max} = 24\,\GeV$. In turn this leads then to a restriction for
the maximal values of the photon virtualities $Q^2_{max}$. The
situation gets a bit more complicated because an effort to obtain
high statistics data leads to a $\nu - Q^2$ correlation and then to
stronger restrictions on $Q^2_{max}$ at different fixed values of
$l_c$. For example, at $\la l_c\ra = 1.35\,\fm$ the value of
$Q^2_{max} = 3.5\,\GeV^2$; and at $\la l_c\ra = 2.45\,\fm$ the value
is $Q^2_{max} = 1.5\,\GeV^2$, as follows from HERMES experiment
\cite{hermes-a2,borissov-06}.

Therefore, we present in Fig.~\ref{hermes} the $Q^2$ dependence of
the nuclear transparency $Tr_A^{inc}$, for exclusive
electroproduction of $\rho$ mesons on $^{207}Pb$ target at different
fixed $l_c$- values. The top and bottom dotted lines represent the
predictions at fixed $l_c = 1.0$ and $5.0\,\fm$, respectively. These
two values limit approximately the range of $l_c$-values
corresponding to a mean $\la\nu\ra = 13\,\GeV$ and $0.5 < Q^2 <
5.0\,\GeV^2$, which follow from the HERMES kinematics
\cite{borissov-06}. Because $l_c\lsim R_{Pb}$, one can expect a
strong onset of CL effects, as one can see in Fig.~\ref{hermes} as
the difference between the top and bottom dotted lines.

Different fixed $l_c$- values determine different maximal photon
virtualities $Q^2_{max}$ allowed for the experimental analysis of
the CT phenomenon. In order to reach the maximal possible values of
$Q^2$ one should fix $l_c$ at its minimal possible values, keeping
reasonable statistics of the data. Such a situation is depicted by
the top dotted line in Fig.~\ref{hermes}, and allows to study the
onset of CT effects up to $Q^2_{max} = 5.0\,\GeV^2$ within the
HERMES kinematics. The higher fixed values of $l_c$ mean a stronger
restriction for $Q^2_{max}$, and consequently a weaker onset of CT
effects. For example, at the second fixed value of $l_c = 5.0\,\fm$
one can study CT effects only up to $Q^2_{max} = 1.3\,\GeV^2$,
although the model calculations shown by the bottom dotted line in
Fig.~\ref{hermes} formally reach the value $Q^2_{max} =
5.0\,\GeV^2$.

As a demonstration of the importance of CL effects in the HERMES
kinematic range we also present, by the solid line in
Fig.~\ref{hermes}, predictions for nuclear transparency at fixed
mean photon energy $\la\nu\ra = 13\,\GeV$. This corresponds to a
change of CL from $5.0$ to $1.0\,\fm$. Because both CL and CT
effects affect the rise of $Tr_A$ with $Q^2$, the net observable
effect is much more visible than that when the effect of CL is
eliminated by the prescription (\ref{534}) (compare each dotted line
with the solid one in Fig.~\ref{hermes}).

\begin{figure}[bht]
\includegraphics{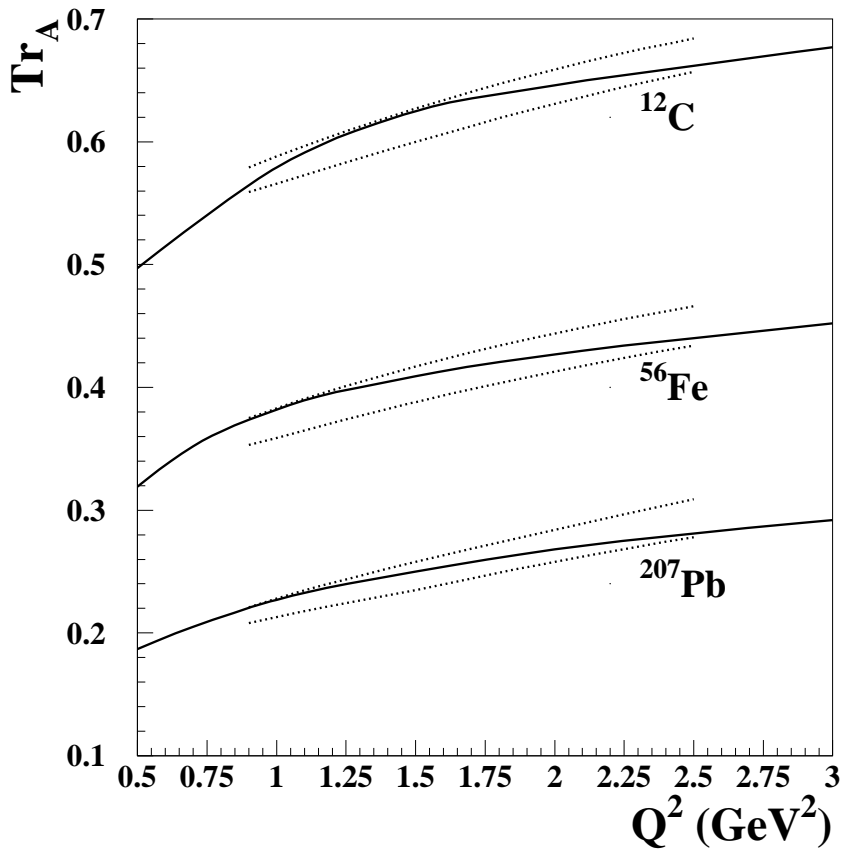}
\begin{center}
\vspace{11.0cm}
\parbox{13cm}
{\caption[Delta]
 {\sl
$Q^2$ dependence of the nuclear transparency
$Tr_A^{inc}$ for exclusive electroproduction of $\rho$
mesons
on nuclear targets $^{12}C$, $^{56}Fe$ and $^{207}Pb$
(from top to bottom).
The top and bottom dotted line with respect to each solid line
represents predictions fixing CL at $l_c = 0.9$
and $0.5\,\fm$, respectively. The solid lines represent predictions
at the fixed mean photon energy $\la\nu\ra = 3.5\,\GeV$
corresponding to CLAS experiment at JLab.
}
 \label{clas}}
 \end{center}
 \end{figure}

Now we switch on to the low energy region corresponding to the CLAS
collaboration at JLab. Because the new data will appear soon we
provide predictions for the expected onset of CT effects for
different nuclei. The model calculations for nuclear transparency as
a function of $Q^2$ are depicted in Fig.~\ref{clas} for several
nuclear targets $^{12}C$, $^{56}Fe$ and $^{207}Pb$. According to the
CLAS kinematics \cite{kawtar}, the values of CL changes from $l_c =
0.9$ to $0.5\,\fm$ when the photon virtuality rises from $Q^2 = 0.9$
to $2.5\,\GeV^2$. Eliminating the CL effects by fixing $l_c = 0.9$
and $0.5\,\fm$, one can obtain the predictions shown in
Fig.~\ref{clas} by the difference between the dotted top and bottom
lines with respect to each solid line, for different nuclear
targets. One can see that in the CLAS kinematic range the CL
effects are rather weak, much weaker than the observed signal of CT.
This comes from the rather small photon energies for values of CL,
$l_c\ll R_A$. A weak onset of CL effects can be seen in
Fig.~\ref{clas} as the difference between the top and bottom dotted
lines, for each nuclear target.

Therefore in contrast with the HERMES kinematics one can study the
variation of nuclear transparency with $Q^2$ at approximately fixed
photon energy, $\nu = 3.5-3.6\,\GeV$, using maximal statistics of
the data. Such a situation is depicted in Fig.~\ref{clas} by the
solid lines, for different nuclear targets. It allows to reach much
higher values of $Q^2$ than those reached in the HERMES experiment,
using the CL-elimination procedure described above.

%
%
%
%

As a consequence of a small onset of CL effects the corresponding
maximal CLAS value of photon virtuality used for studying CT effects
$Q^2_{max}\sim 2.5-3.0\,\GeV^2$ does not differ much from the HERMES
value $Q^2_{max}\sim 3.5-4.0\,\GeV^2$, although the mean photon
energy is four times lower than that in the HERMES experiment. It
means that the CLAS kinematic range leads to a more effective
investigation of CT phenomenon than in a case when we are forced to
use the CL-elimination procedure, i.e. keeping $l_c = constant$ at
different values of $Q^2$, which is typical for the HERMES
experiment.

Such an advantage in the investigation of CT effects in the CLAS
experiment should be maintained also after its future upgrade to a
beam energy of $12\,\GeV$ \cite{will}. As a consequence of larger
photon energies, larger values of the photon virtualities $Q^2$ will
be reached, keeping the condition of a weak onset of CL effects
$l_c\ll R_A$ for heavy nuclear targets. Thus it allows to study a
stronger onset of CT effects investigating nuclear transparency at
larger values of $Q^2$.

%
\section{Summary and conclusions}\label{conclusions}
%

Electroproduction of vector mesons off nuclei is a very effective tool
for studying the interplay between coherence (shadowing) and
formation (color transparency) effects. In the present paper we
investigated how those effects manifest themselves in the production
of vector mesons off nuclei, at different energies corresponding to
the E665, HERMES and CLAS experiments. We used, from \cite{knst-01},
a rigorous quantum-mechanical approach based on the light-cone QCD
Green function formalism, which naturally incorporates these
interference effects. Motivated by expected new data from the CLAS
collaboration we studied the onset of CT effects at small energies,
predicting a rising nuclear transparency as a function of $Q^2$.

As the first step we discussed CT effects at large energies
corresponding to E665 experiment. It was already shown in
Ref.~\cite{knst-01} that the high energy limit ($l_f\sim l_c\gg
R_A$) can be applied only at small and medium values of photon
virtualities, $Q^2\lsim 3\div 5\,\GeV^2$. Here the expressions for
nuclear production cross sections are sufficiently simplified. This
so called ``frozen'' approximation includes only CT, because there
are no fluctuations of the transverse size of the $\bar qq$ pair. At
larger values of $Q^2\gsim 5\,\GeV^2$ both the CL $l_c\lsim R_A$ and
the onset of CL effects start to be important. The ``frozen''
approximation cannot be applied anymore for the study of a signal of
CT effects. Therefore one should use the general expression
Eq.~(\ref{520}), which incorporates in addition CL effects and
therefore interpolates between the previously known low and high
energy limits for incoherent production of vector mesons.

In the incoherent electroproduction of vector mesons at low and
medium energies the onset of coherence effects (shadowing) can mimic
the expected signal of CT. Both effects, CT and CL, work in the same
direction. In order to single out the formation effect the
experimental data should be binned in $l_c$ and $Q^2$
\cite{knst-01}. The observation of a rising nuclear transparency as
function of $Q^2$ for fixed $l_c$ would signal CT. Such a procedure
for the elimination of CL effects must be applied at medium
energies, corresponding to the HERMES kinematic range, when
$l_c\lsim R_A$. We showed that a $Q^2$-variation of CL affects a
strong onset of CL effects as one can see in Fig.~\ref{hermes} as a
difference between the top and bottom dotted lines.

At still smaller photon energies, corresponding to the CLAS
experiment, both $l_c$ and $l_f$ are short enough compared to the
mean spacing of the bound nucleons. Consequently, the CL effects are
rather weak, and in fact much weaker than the observed effect of CT
(see the difference between the top and bottom dotted lines for each
nuclear target in Fig.~\ref{clas}). This provides an advantage in
the investigation of the onset of CT when one can study the rise of
nuclear transparency with $Q^2$ at fixed photon energies, processing
the data at maximal statistics (see the solid lines in
Fig.~\ref{clas}). We showed that such a procedure allows to reach
much higher values of $Q^2$ effective for studying the CT
phenomenon, than those used in the HERMES experiment applying the
CL-elimination prescription of fixing $l_c$ (see Eq.~\ref{534}).

Concluding, the exploratory study of CT effects at different
energies in electroproduction of light vector mesons off nuclei
opens new possibilities for the investigation of different
manifestations of CL effects. It gives a good basis for the further
effective studies of the onset of CT effects, after an upgrade of
the CLAS experiment at JLab to a $12\,\GeV$ energy electron beam.

\bigskip

{\bf{Acknowledgments:}}
 This work was supported in part by Fondecyt (Chile) grant 1050519,
by DFG (Germany)  grant PI182/3-1, and by the Slovak Funding
Agency, Grant No. 2/7058/27.

 \def\appendix{\par
 \setcounter{section}{0} \setcounter{subsection}{0}
 \def\thesection{Appendix \Alph{section}}
 \def\thesubsection{\Alph{section}.\arabic{subsection}}
 \def\theequation{\Alph{section}.\arabic{equation}}
 \setcounter{equation}{0}}

\end{document}